\documentclass[twocolumn,submitted]{IEEEtran}
\usepackage[left=0.7 in, right=0.5 in, top=0.7 in, bottom=0.7 in]{geometry} % Document margins

\usepackage{hyperref}
\usepackage{graphicx}
\usepackage{cite}
\usepackage{amsmath}
\usepackage{color}
\begin{document}
\title{\bf An Investigation of Magnetic Hysteresis Error in Kibble Balances}
\author{%BIPM Kibble balance group, V6
\thanks{Manuscript of IEEE Trans. Instrum. Meas. All authors are with the Bureau International des Poids et Mesures (BIPM), 92312 S\`{e}vres, France. Email: leeshisong@sina.com; fang@bipm.org.}
%\thanks{Manuscript version R2, latest updated in 12/2019.}
Shisong Li, {\it Senior Member, IEEE}, Franck Bielsa, Michael Stock, Adrien Kiss and Hao Fang
%Email: leeshisong@sina.com; fang@bipm.org
}
\maketitle

\begin{abstract}
Yoke-based permanent magnetic circuits are widely used in Kibble balance experiments. In these magnetic systems, the coil current, with positive and negative signs in two steps of the weighing measurement, can cause an additional magnetic flux in the circuit and hence a magnetic field change at the coil position. The magnetic field change due to the coil current and related systematic effects have been studied with the assumption that the yoke material does not contain any magnetic hysteresis. In this paper, we present an explanation of the magnetic hysteresis error in Kibble balance measurements. An evaluation technique based on measuring yoke minor hysteresis loops is proposed to estimate the effect. The dependence of the magnetic hysteresis effect and some possible optimizations for suppressing this effect are discussed.
\end{abstract}

\begin{IEEEkeywords}
Kibble balance, magnetic circuit, $BH$ hysteresis loop, measurement error.
\end{IEEEkeywords}
\IEEEpeerreviewmaketitle
\pagenumbering{gobble}
%\clearpage
%{\color{blue}
\section{Introduction}
The Kibble balance, formerly known as the watt balance \cite{kb76}, is one of the major instruments for realizing the unit of mass, i.e. the kilogram, in terms of the Planck constant $h$ under the revised International System of Units (SI) \cite{cgpm18}. A Kibble balance establishes a relationship between the mass of an artefact and the Planck constant by comparing mechanical power to electrical power, which can be viewed as a bridge linking classical and quantum mechanics \cite{darine16}. Detailed principles and descriptions of a Kibble balance can be found in recent review papers, e.g. \cite{stephan16}.

In Kibble balance experiments, a sub-Tesla magnetic field with a good vertical uniformity is required. After many years of optimization and practice, all the ongoing Kibble balances in the world have chosen to use yoke-based permanent magnetic circuits \cite{4,5,6,7,8,9,10,11,12,13}. One of the main effects of such a magnet system is the coil-current effect, i.e. the coil current interacts with the main magnetic circuit and can cause a change in the magnetic field at the measurement position.

Theoretical and experimental studies have been made on the current effect: In \cite{li1,li2}, the static nonlinear effect was studied, and the main static non-linearity was found to be due to the yoke magnetic status change in the weighing measurement. The effect was evaluated to be small compared to a typical Kibble balance measurement uncertainty, e.g. $2\times10^{-8}$. The nonlinear current effect can be precisely determined by running the experiment with different currents (masses). Experimental measurements of Kibble balances at the National Research Council (NRC, Canada) and the National Institute of Standards and Technology (NIST, USA) yielded a nonlinear current term with a few parts in $10^9$ in their systems \cite{16,17}.
In \cite{18}, it was shown that the linear current effect is mainly contributed by the coil inductance change at different vertical positions, and a significant linear magnetic profile change, proportional to the coil current, has been experimentally observed. Further, different profile changes in weighing and velocity phases were experimentally measured in one-mode schemes \cite{19}. The linear change of the magnetic profile can cause a bias when there is a coil vertical position change during two steps of weighing, i.e. mass-on and mass-off. This bias, which is also closely related to parameters of the magnet system, in general, should be carefully considered in the measurement.

So far, all studies of the current effect are made based on an assumption that the yoke has a fixed magnetization curve and does not contain any hysteresis. In reality, the yoke used to build the Kibble balance magnet, as reported in \cite{21}, has a considerable magnetic hysteresis. The excitation (current) change during the weighing measurement can shift the yoke working point and hence could bring a magnetic field change at the coil position compared to the field measured in the velocity phase. In this paper, theoretical analysis and experimental investigations are presented to build an evaluation technique for potential errors caused by the magnetic hysteresis.
The analysis assumes that the magnet yoke is left in a magnetized state at the end of the weighing phase. We note this magnetization state is not unavoidable. It can practically be erased by applying a decaying oscillatory waveform to the coil current at the end of the weighing phase, as implemented in the NPL (National Physical Laboratory, UK) and NRC Kibble balances \cite{4,5}. The remainder of the paper is organized as follows: In section \ref{sec02}, a theoretical analysis of the magnetic hysteresis effect is presented. In section \ref{sec03}, an experimental example is taken for an estimation of the magnetic hysteresis error. Some discussions on the dependence of the effect, as well as possible ways to suppress the hysteresis error, are summarized in section \ref{sec04}.

\section{Theoretical analysis}
\label{sec02}
\subsection{Overview of the analysis}
The main purpose of this paper is to establish the relationship of magnetic flux density change at the coil position, $\Delta B_a$, due to the coil current $I$ and the yoke $BH$ hysteresis. The analysis begins with a conventional two-mode, two-phase measurement scheme, and is divided into three independent steps:

\begin{enumerate}
  \item Linking the magnetic flux density change at the coil position $\Delta B_a$ to the magnetic field change in the yoke $\Delta B_y$, i.e. function $\Delta B_a=\mathcal{A}_1(\Delta B_y)$.
  \item Modeling the magnetic flux density change of the yoke $\Delta B_y$, via yoke $BH$ minor hysteresis loops, as a function of the magnetic field change in the yoke $\Delta H_y$. i.e. $\Delta B_y=\mathcal{A}_2(\Delta H_y)$.
  \item Expressing the magnetic field change in the yoke, $\Delta H_y$, as a function of the coil current in the weighing phase, i.e. $\Delta H_y=\mathcal{A}_3(I)$.
\end{enumerate}
Using the three steps above, the magnetic field change at the coil position can be modeled in terms of coil current in the weighing measurement, and the magnetic hysteresis effect can be evaluated accordingly.

\subsection{Linking $\Delta B_a$ to $\Delta B_y$}
\begin{figure}
\center
\includegraphics[width=0.5\textwidth]{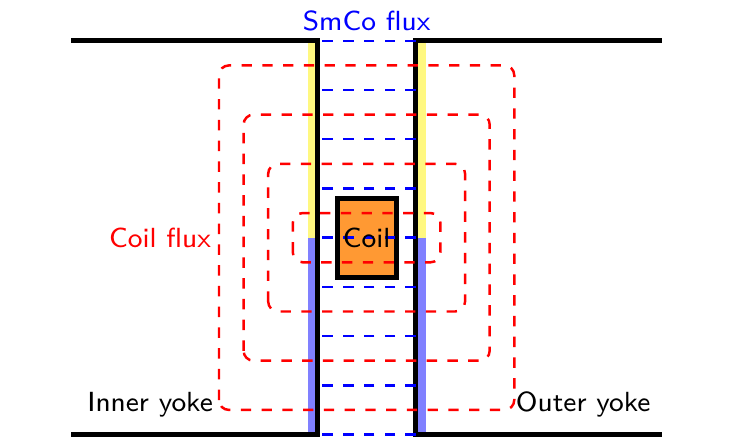}
\caption{The magnetic flux distribution in the air gap region. The blue and red dashed lines denote respectively the SmCo flux and the coil flux. Different colors (yellow, blue) marked in the yoke-air boundaries denote opposite signs of magnetic flux density change.}
\label{fig01}
\end{figure}

In Kibble balances, the general idea of a yoke-based magnetic circuit, e.g. the BIPM-type magnet \cite{21}, is to compress the magneto-motive force (mmf) in a narrow air gap formed by inner and outer yokes so that the magnetic field generated is strong and uniform. In the weighing measurement, the coil is placed in the air gap with a current, and the newly created magnetic flux of the coil remains within the path composed of the air gap (goes through twice) and inner/outer yokes. Fig. \ref{fig01} presents the magnetic flux distribution created by the SmCo magnets and the coil current in the air gap region. The total magnetic field will increase when the coil flux has the same direction as the SmCo flux. Otherwise, when two flux directions are opposed, the magnetic field in the air gap is reduced by the coil flux.

In a Kibble balance, the weighing position is usually set at an extreme point (which has the flattest profile, i.e. $\partial B_{ar}/\partial z=0$, where $B_{ar}$ is the radial magnetic flux density in the air gap, and $z$ is the vertical axis) in order to minimize related systematic errors, e.g. the magnetic field difference due to the current asymmetry of mass-on and mass-off \cite{19}. Note that in an up-down symmetrical magnetic circuit design, the field extreme point locates at the vertical center of the air gap, i.e. $z=0$, because, in this plane, the SmCo magnetic flux contains a purely horizontal component, i.e. $B_{az}=0$. Note that in reality, the magnetic center could be shifted by non-ideal construction of the magnet, e.g. the asymmetry of the magnetization, mechanical assembly, etc, but the hysteresis error, which is the focus of this paper has a weak dependence on the field flat point in the central region and can be similarly analyzed. Without losing generality, in the following analysis, we assume that the weighing position is $z=0$, and in this case (as shown in Fig. \ref{fig01}) the coil flux has a symmetrical up-down distribution in both inner and outer yokes. Since only the horizontal component of the magnetic flux density can generate a vertical force, in the following discussion, the magnetic flux density denotes simply the horizontal component. Also, in the following analysis, we take a typical cycled Kibble balance measurement sequence, $VW_1W_2V$($V$ denotes the velocity measurement, $W_1$ and $W_2$ the weighing measurements with different current polarities), as an example. Other measurement sequences can be similar analyzed.

In the velocity measurement, there is no current in the coil and the magnetic flux density at the coil position is $B_{av}$. The $BH$ working points at inner/outer yoke boundaries are respectively, $(H_{yiv}, B_{yiv})$ and $(H_{yov}, B_{yov})$. In the first weighing step, i.e. $I=I_+$ (mass-on), the flux produced by the current shifts the magnetic flux density of inner and outer yokes to $B_{yi+}$ and $B_{yo+}$, and in the second step $I=I_-$ (mass-off), the magnetic flux density in the inner and outer yokes is changed to $B_{yi-}$ and $B_{yo-}$.

It is known that in such a magnetic circuit, the horizontal magnetic flux density in the air gap follows approximately a $1/r$ relationship, where $r$ is the radius of the focused position in the air gap \cite{li16}. In a typical Kibble balance magnet, the air gap width is usually much smaller than its radius, and hence the coil field gradient in the air gap can be approximately considered linear along $r$ direction. This approximation allows us to write the magnetic field at the coil position as a function of the inner and outer yoke boundaries as
\begin{equation}
\Delta B_a\approx\frac{\Delta B_{yi}+\Delta B_{yo}}{2},
\label{eq1}
\end{equation}
where $\Delta B_{yi}$ and $\Delta B_{yo}$ denote the magnetic flux density change respectively at the inner and outer yoke boundaries. Note that (\ref{eq1}) is not accurate for absolute field calculation, but it is good enough for modeling the field change where the total effect is small (verified in section \ref{subD}).

Since the normal component of magnetic flux density is continuous across the yoke-air interface, the magnetic flux density at the coil position in two steps of the weighing measurement can be written as
\begin{eqnarray}
  B_{a+} &=& B_{av}+\frac{(B_{yi+}-B_{yiv})+(B_{yo+}-B_{yov})}{2} \nonumber\\
  B_{a-} &=& B_{av}+\frac{(B_{yi-}-B_{yiv})+(B_{yo-}-B_{yov})}{2}
  \label{eq2}
\end{eqnarray}
Eq. (\ref{eq2}), by combining with the weighing equations,
\begin{eqnarray}
  B_{a+}I_+ &=& \frac{mg-m_cg}{L} \nonumber\\
  B_{a-}I_- &=& \frac{-m_cg}{L}
  \label{eq3}
\end{eqnarray}
where $L$ denotes the coil wire length, $m$ the mass of the artefact, $m_c$ the counter mass (including the reading of the weighing cell or a mass comparator), $g$ the local gravitational acceleration, gives the effective magnetic field density seen by the coil in the weighing phase as
\begin{eqnarray}
B_{aw}&=&\frac{mg}{(I_+-I_-)L}\nonumber\\
&=&B_{av}+\frac{(B_{yi+}-B_{yiv})I_+-(B_{yi-}-B_{yiv})I_-}{2(I_+-I_-)}\nonumber\\
&+&\frac{(B_{yo+}-B_{yov})I_+-(B_{yo-}-B_{yov})I_-}{2(I_+-I_-)}.
\label{eq4}
\end{eqnarray}
Conventionally, as the currents in the weighing phase are set symmetrically, i.e. $I_+=-I_-=I$, then (\ref{eq4}) can be simplified as
\begin{eqnarray}
  \frac{B_{aw}-B_{av}}{B_{av}}&=&\frac{(B_{yi+}-B_{yiv})+(B_{yi-}-B_{yiv})}{4B_{av}}\nonumber\\
  &+&\frac{(B_{yo+}-B_{yov})+(B_{yo-}-B_{yov})}{4B_{av}}
  \label{eq5}
\end{eqnarray}
Note that in Kibble balances, only the magnetic field change between two measurement phases, as presented in (\ref{eq5}), can introduce a measurement error. Eq. (\ref{eq5}) links the magnetic flux density change at the coil position to the magnetic flux density change at the inner/outer yoke boundaries. When the yoke field change is a linear function of the current, i.e. $B_{yi+}-B_{yiv}=-(B_{yi-}-B_{yiv})$, $B_{yo+}-B_{yov}=-(B_{yo-}-B_{yov})$, the magnetic field change at the coil position is averaged out. However, the yoke magnetic flux density change as a function of the coil current (or $H$ field change) is not linear (even without hysteresis), and the residual nonlinear term should be evaluated.

\subsection{$\Delta B_y(\Delta H_y)$ and minor $BH$ hysteresis loops}
\begin{figure}
\center
\includegraphics[width=0.5\textwidth]{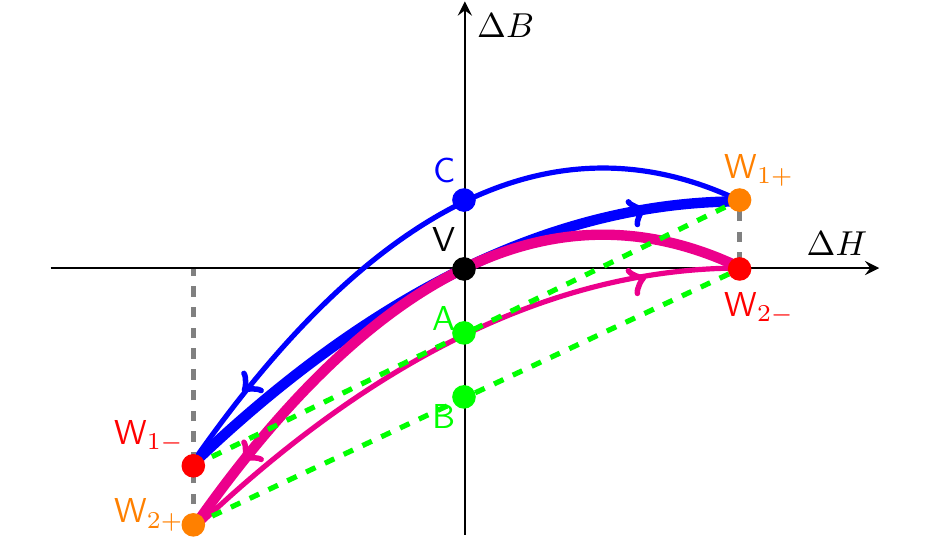}\\(a)\\
\includegraphics[width=0.5\textwidth]{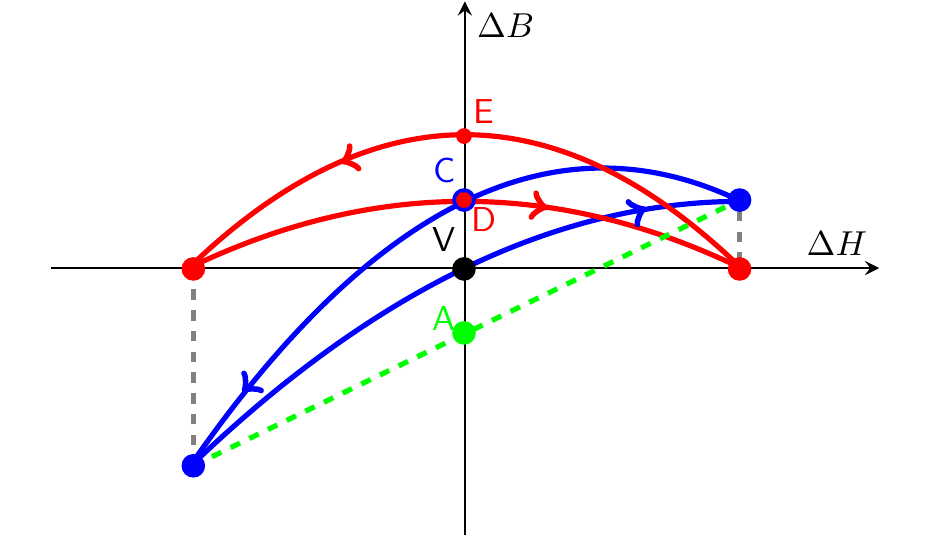}\\(b)
\caption{Magnetic status change of the yoke in the weighing phase. (a) shows the yoke $BH$ working point change due to the coil flux during mass-on and mass-off. In (b), the linear component of the minor loop is removed, yielding a flat normalized minor loop. Note that $\Delta B_y$ and $\Delta H_y$ are defined as the magnetic flux density and magnetic field change compared to the yoke status in the velocity measurement, i.e. $\Delta B_y=B_y-B_{yv}$, $\Delta H_y=H_y-H_{yv}$.}
\label{fig02}
\end{figure}

In the Kibble balance magnet, a good design would keep the yoke permeability, $\mu$, at or close to the maximum point of the $\mu(H)$ curve, where the $B(H)$ curve has a large static slope $B/H$. When the yoke magnetic field $H$ is slightly shifted by the coil flux in the weighing phase, a minor $BH$ loop, $(H_{yv},B_{yv})$- $(H_{y+},B_{y+})$- $(H_{y-},B_{y-})$- $(H_{yv},B_{yv})$, is formed.
Fig. \ref{fig02}(a) presents an example of the minor $BH$ hysteresis loop in the weighing measurement. The shape of the minor loop can be different but the global slope of the loop is always positive (determined by the differential permeability $\mu_d$). For either the inner or outer yoke, the coil flux with mass-on will increase the horizontal magnetic field in one half of the yoke boundary, while it will decrease the field of the other half of the yoke boundary by the same amount.  For example, the magnetic field in the upper half of the yoke boundary is shifted by $\Delta \mathcal{H}$, while the field of the lower half is shifted by $-\Delta \mathcal{H}$. As shown in Fig. \ref{fig02}(a), with plus current (mass-on), the $BH$ working point of the upper yoke is shifted to $W_{1+}$ following the blue minor loop, and the lower yoke $BH$ working point moves to $W_{2+}$ along the magenta loop. With the negative current (mass-off), the upper yoke is working at $W_{1-}$ and the lower yoke would be at $W_{2-}$.

Based on (\ref{eq5}), the average magnetic flux density change for the inner yoke can be written as
\begin{eqnarray}
  &&(B_{yi+}-B_{yiv})+(B_{yi-}-B_{yiv})\nonumber \\
  &=&\left(\frac{B_{yi1+}+B_{yi2+}}{2}-B_{yiv}\right)+\left(\frac{B_{yi1-}+B_{yi2-}}{2}-B_{yiv}\right)\nonumber\\ &=&\left(\frac{B_{yi1+}+B_{yi1-}}{2}-B_{yiv}\right)+\left(\frac{B_{yi2+}+B_{yi2-}}{2}-B_{yiv}\right).
  \label{eq6}
\end{eqnarray}
On the right side of (\ref{eq6}), the first term denotes the nonlinearity of $H$ increasing curve ($dH/dt>0$) and the second term presents the nonlinearity of $H$ decreasing curve ($dH/dt<0$). $(B_{yi1+}+B_{yi1-})/2$ is the averaged magnetic flux density of $W_{1+}$ and $W_{1-}$, which equals the magnetic flux density at point A. $(B_{yi2+}+B_{yi2-})/2$ is the magnetic flux density at point B.
Then (\ref{eq6}) can be rewritten as
\begin{equation}
(B_{yi+}-B_{yiv})+(B_{yi-}-B_{yiv})=-(\overline{AV}+\overline{BV}),
\label{eq7}
\end{equation}
where $\overline{AV}$ and $\overline{BV}$ denote the line lengths of $AV$ and $BV$. Since the two minor loops in Fig. \ref{fig02}(a) have the same shape, and hence $\overline{BV}=\overline{AC}$. Then (\ref{eq7}) is simplified to
\begin{equation}
(B_{yi+}-B_{yiv})+(B_{yi-}-B_{yiv})=-(\overline{AV}+\overline{AC}).
\label{eq8}
\end{equation}
Eq. (\ref{eq8}) can be simplified by normalizing the hysteresis curves as shown in Fig. \ref{fig02}(b), i.e. two end points of the original minor loop are rotated to be aligned with the $\Delta H_y$ axis. It can be mathematically proven (as shown in Appendix) that
\begin{equation}
-(\overline{AV}+\overline{AC})=-(\overline{VD}+\overline{VE})=\mathcal{F}(\Delta \mathcal{H}^2),
\label{eq9}
\end{equation}
where $\mathcal{F}$ denotes a function related only to the even-order terms of yoke magnetic field change due to the coil current $\Delta \mathcal{H}$.
When the minor loop is flat, we can rewrite (\ref{eq6}) as
\begin{eqnarray}
&&(B_{yi+}-B_{yiv})+(B_{yi-}-B_{yiv})\nonumber\\
&=&-\Delta \mathcal{B}_i|_{dH/dt>0}-\Delta \mathcal{B}_i|_{dH/dt<0},
\label{eq10}
\end{eqnarray}
where $\Delta \mathcal{B}_i|_{dH/dt<0}$, $\Delta \mathcal{B}_i|_{dH/dt>0}$, as shown in Fig. \ref{fig02}(b), denote the magnetic flux density change on normalized $H$ decreasing and $H$ increasing curves at $\Delta H_y=0$, i.e. the magnetic flux density at points D and E respectively.

A similar analysis can be applied to the outer yoke. The magnetic flux density change in the yoke boundary due to the magnetic hysteresis is,
\begin{eqnarray}
&&(B_{yo+}-B_{yov})+(B_{yo-}-B_{yov})\nonumber \\
&=&-\Delta \mathcal{B}_o|_{dH/dt>0}-\Delta \mathcal{B}_o|_{dH/dt<0}.
\label{eq11}
\end{eqnarray}
Substituting (\ref{eq10}) and (\ref{eq11}) into (\ref{eq5}) yields
\begin{eqnarray}
  &&\frac{B_{aw}-B_{av}}{B_{av}}\nonumber\\
  &=&-\frac{\Delta \mathcal{B}_i|_{dH/dt>0}+\Delta \mathcal{B}_i|_{dH/dt<0}}{4B_{av}}\nonumber\\
  &&-\frac{\Delta \mathcal{B}_o|_{dH/dt>0}+\Delta \mathcal{B}_o|_{dH/dt<0}}{4B_{av}}
  \label{eq12}
\end{eqnarray}
Since the $BH$ working point of inner and outer yoke boundaries is not far separated (magnetic flux density difference $<0.1$\,T), (\ref{eq12}) can be approximated as
\begin{eqnarray}
  &&\frac{B_{aw}-B_{av}}{B_{av}}\nonumber\\
  &=&-\frac{\Delta \mathcal{B}|_{dH/dt>0}+\Delta \mathcal{B}|_{dH/dt<0}}{2B_{av}},
  \label{eq1x}
\end{eqnarray}
where $\Delta \mathcal{B}$ is the yoke magnetic flux density change (normalized curve) with $B_y\approx(B_{yi}+B_{yo})/2$. In order to evaluate the hysteresis effect, the minor $BH$ loops and the yoke magnetic field change due to the current, i.e. $\Delta H$, should be known.

\subsection{Yoke magnetic field change due to coil current}
\label{subD}
The magnetic profile change due to the coil flux in the weighing phase has been studied in \cite{19}. The slope of the magnetic profile change can be measured directly by the force-current ratio at different positions or by the voltage-velocity ratio with a moving current-carrying coil. To evaluate the hysteresis effect, the real magnetic field distribution along the vertical direction is required. A study of the field change based on the finite element analysis (FEA) is presented \cite{22}.

\begin{figure}
\center
\includegraphics[width=0.5\textwidth]{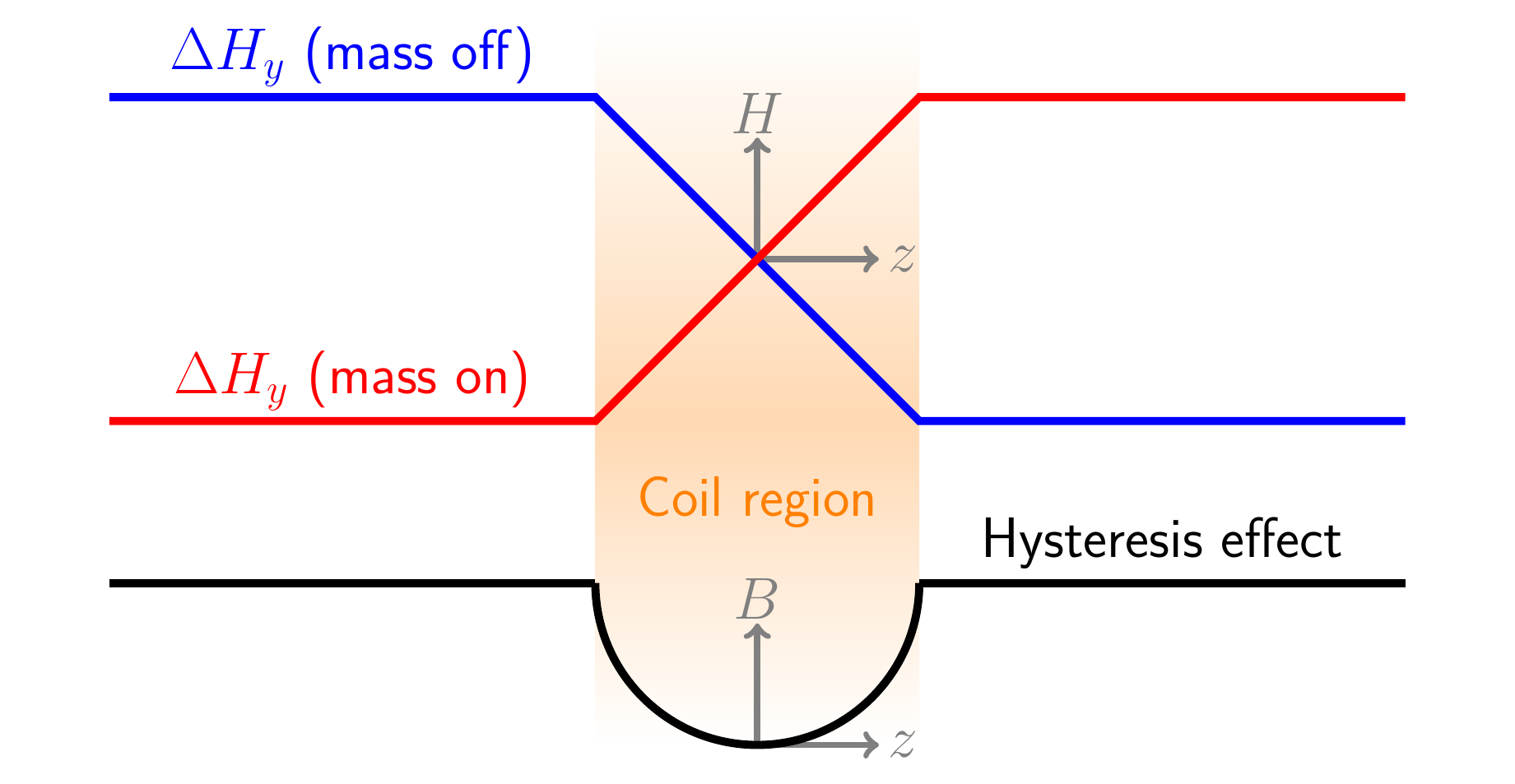}
\caption{The magnetic field change due to the coil current and its related magnetic hysteresis effect. At the coil center $z=0$\,mm, the $\Delta H_y$ value is zero and the magnetic hysteresis error at this point is also zero. Above or below the coil, both $|\Delta H_y|=\Delta\mathcal{H}$ and the hysteresis effect reach maxima. The hysteresis effect seen by the coil should be the average in the coil region.}
\label{fig03}
\end{figure}

As shown in Fig. \ref{fig03}, the magnetic flux density change caused by the coil current is a step function along the vertical direction: At two ends, the field change stays constant (opposite directions) and in the coil region, the magnetic field change is a linear function of the coil vertical position $z$. This result can be easily modeled by Ampere's law: Above or below the coil, the ampere-turns of the coil, i.e. $NI$ ($N$ is the total number of the coil winding), is fixed. In the coil region, the ampere-turns are $NIz/h_c$, where $h_c$ is the half height of the coil. The main mmf drop is horizontally along the air gap (twice), then the magnetic flux density produced by the coil current at the air gap center $r=r_c$ can be calculated as
\begin{eqnarray}
\Delta B=\left\{
\begin{matrix}
  \displaystyle\frac{\mu_0NI}{2\delta}\mbox{sign}(z),~~|z|\geq h_c\\
  ~\\
  \displaystyle\frac{\mu_0NIz}{2\delta h_c},~~|z|<h_c
\end{matrix}
\right.
\label{eq13}
\end{eqnarray}
where $\delta$ is the width of the air gap, $\mu_0$ the permeability of vacuum, sign$(z)$ the sign of $z$. This calculation is checked using an FEA example. In the simulation, an air gap with similar parameters of the BIPM Kibble balance magnet is used: The radii of the inner and outer yoke boundaries are $r_i=118.5$\,mm and $r_o=131.5$\,mm. The height of the air gap is 80\,mm. The air gap width is $\delta=13$\,mm. The coil, 10\,mm in width and 20\,mm in height ($h_c=10$\,mm), is placed at the geometrical center of the air gap, i.e. $r_c=125$\,mm and $z=0$\,mm. The ampere-turns of the coil, $NI$, are 14\,A, which can generate 4.9\,N magnetic force (corresponding to the weight of a 500\,g mass).

\begin{figure}
\center
\includegraphics[width=0.5\textwidth]{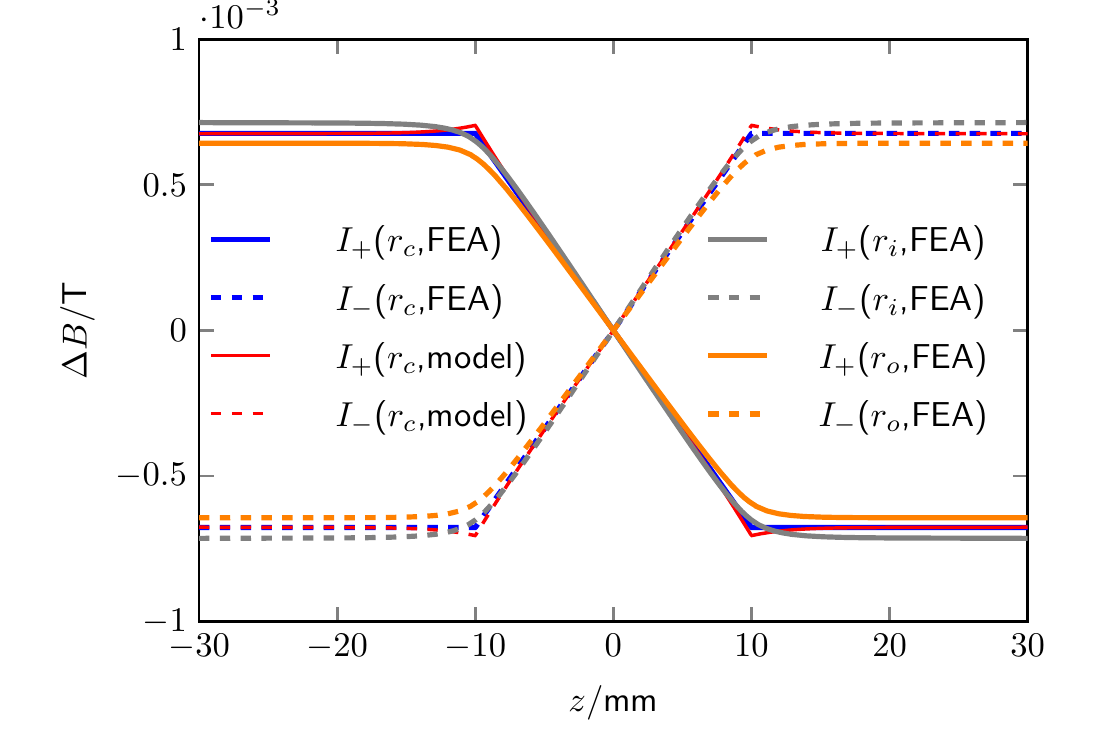}
\caption{The coil magnetic flux density distribution along the vertical direction at different radii.}
\label{fig04}
\end{figure}

Fig. \ref{fig04} compares the magnetic flux density distribution obtained by FEA and an analytical model (\ref{eq13}). The result produced by the analytical model agrees well with the FEA calculation. Fig. \ref{fig04} also presents the magnetic flux density distribution at the inner and outer yoke boundaries, $r_i$ and $r_o$. It can be observed that for larger radius, the magnetic flux density is lower. The maximum magnetic flux density (above or below the coil) is 0.7138\,mT, 0.6766\,mT and 0.6431\,mT respectively at $r_i=118.5$\,mm, $r_c=125$\,mm and $r_o=131.5$\,mm. These values confirm the $1/r$ distribution of the coil magnetic flux density in the air gap. The result presented in Fig. \ref{fig04} provides a good check on the approximation given in (\ref{eq1}). The difference between the result obtained by (\ref{eq1}) and the $1/r$ relationship (real distribution) is only 0.3\%.

Knowing the $\Delta B$ value in the air gap center $r_c$ based on (\ref{eq13}), we can then solve the magnetic flux density at both yoke-air boundaries following the $1/r$ relationship. Since the horizontal component of the magnetic flux density is continuous in both yoke-air boundaries, the magnetic field changes at the inner and outer yokes are solved respectively as
\begin{eqnarray}
\Delta H_{yi}=\left\{
\begin{matrix}
  \displaystyle\frac{NIr_c}{2\mu_rr_i\delta }\mbox{sign}(z),~~|z|\geq h_c\\
  ~\\
  \displaystyle\frac{NIr_cz}{2\mu_rr_ih_c\delta },~~|z|<h_c
\end{matrix}
\right.
\label{eq14}
\end{eqnarray}

\begin{eqnarray}
\Delta H_{yo}=\left\{
\begin{matrix}
  \displaystyle\frac{NIr_c}{2\mu_rr_o\delta }\mbox{sign}(z),~~|z|\geq h_c\\
  ~\\
  \displaystyle\frac{NIr_cz}{2\mu_rr_oh_c\delta},~~|z|<h_c
\end{matrix}
\right.
\label{eq15}
\end{eqnarray}
where $\mu_r$ is the relative yoke permeability. As shown in Fig. \ref{fig03}, at the coil center $z=0$, $\Delta H_y$ is zero, and the magnetic hysteresis effect at this point is also zero. It is shown in (\ref{eq9}) that the magnetic hysteresis effect is related only to the even order of yoke $H$ field change. Therefore, at $z=\pm h_c$, both $|\Delta H_y|$ and the magnetic hysteresis effect reach a maxima. Note, we define the maximum magnetic field change at yoke boundary as $\Delta\mathcal{H}$. As a result, the hysteresis effect seen by the measurement should be the average value in the coil region. If the magnetic flux density change in the yoke $\Delta\mathcal{B}$ is described by polynomial forms of $\Delta \mathcal{H}^2$ as
\begin{equation}
\Delta\mathcal{B}=\sum_{i=2,4,...}\kappa_i\Delta \mathcal{H}^i,
\label{eq16}
\end{equation}
a following gain factor should be added to the maximum effect (at $z=\pm h_c$) due to the average in the coil range, i.e.
\begin{equation}
K=\int_{0}^{1}\sum_{i=2,4,...}\frac{\kappa_ix^i}{\sum_{i=2,4,...}\kappa_i} dx=\frac{\sum_{i=2,4,...}\kappa_i/(i+1)}{\sum_{i=2,4,...}\kappa_i},
\label{eq16x}
\end{equation}
where $x$ is the normalized hysteresis effect ranged from 0 to 1. Note that the gain factor $K$ is equal to 1/3 when $\Delta\mathcal{B}$ is described only by the quadratic term, i.e. $i=2$. Another conclusion obtained from the above analysis is that the magnetic hysteresis does not rely on the coil height ($2h_c$), because the maximum effect value and $K$ are both independent of $h_c$.

\subsection{Measurement and evaluation technique}
Using the above analysis, the measurement and evaluation technique for the magnetic hysteresis effect is proposed as follows. First, the yoke minor hysteresis loops, centered to the yoke $BH$ working point (air-yoke boundaries or an averaged magnetic flux density close to $B_{yv}$), need to be measured. The purpose of this step is to determine the coefficient $\kappa_i$ in (\ref{eq16}). In Kibble balances, since the yoke $H$ field change due to the current is tiny and cannot be measured directly, here a fitting method is suggested: 1) measure a group of minor hysteresis loops, $H$ centered to $H_{yv}$ and $\Delta \mathcal{H}$ changing as a variable; 2) normalize the hysteresis loops measured; 3) fit $\overline{VD}=\Delta\mathcal{B}|_{dH/dt>0}$ and $\overline{VE}=\Delta\mathcal{B}|_{dH/dt<0}$ as functions of $\Delta \mathcal{H}$; 4) find a best-fit order and calculate $\kappa_i$ values for both $H$ increasing and $H$ decreasing curves. In this way, both $\Delta\mathcal{B}(\Delta\mathcal{H}^2)$ and the gain factor $K$, i.e. (\ref{eq16}) and (\ref{eq16x}), are solved.

The next step is to follow (\ref{eq14}) and (\ref{eq15}) and calculate the $H$ field change at yoke boundaries during weighing measurements. Note that in this step the permeability of the yoke needs to be measured. Knowing $\Delta H_y(z)$ in the coil region during mass-on and mass-off, the yoke magnetic flux density change $\Delta\mathcal{B}$ in both directions can be calculated based on (\ref{eq16}). Then using (\ref{eq12}) or (\ref{eq1x}), the magnetic hysteresis effect can be evaluated.

\section{Experimental measurement: An example}
\label{sec03}
\subsection{Experimental setup}
In this section, we give an estimation of the magnetic hysteresis effect based on an experimental measurement of yoke minor loops. The yoke material used is a Ni-Fe alloy (50:50). The sample has a inner diameter of $r_1=55$\,mm and a outer diameter $r_2=70$\,mm. The thickness of the yoke ring is 15\,mm.

\begin{figure}
\center
\includegraphics[width=0.5\textwidth]{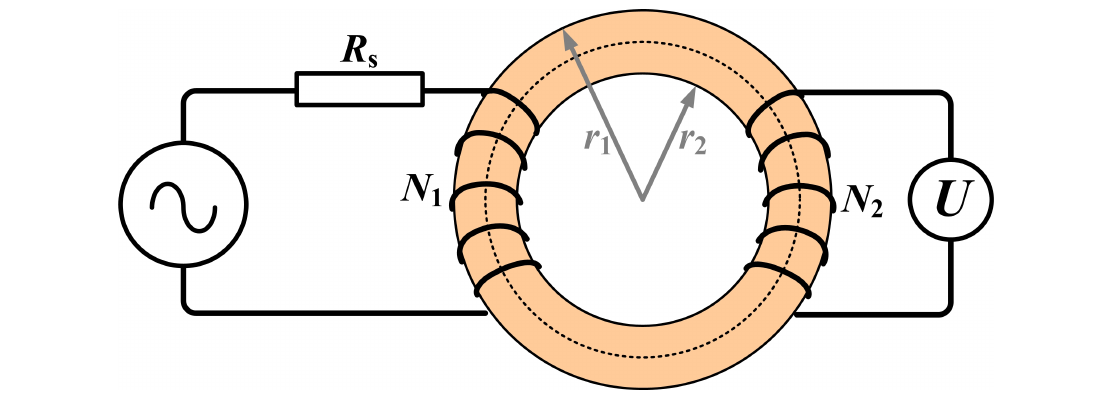}
\caption{Electrical circuit for measuring main and minor hysteresis loops in the yoke material.}
\label{fig05}
\end{figure}

The measurement circuit is presented in Fig. \ref{fig05}. A signal generator, which can output up to 100\,mA current, is used to supply the required current. In order to reduce the influence of eddy current and skin effect, the frequency of the signal used in the measurement is set at 0.1\,Hz. The primary winding, with a total number of turns $N_1$, is excited by the signal generator. The current through the primary winding is measured by the voltage drop on a standard resistor, $R_s=25\Omega$. According to Ampere's law, the magnetic field through the core (yoke) is calculated as
\begin{equation}
H=\frac{N_1\mathcal{I}}{\pi(r_1+r_2)}
\label{eq4.1}
\end{equation}
The induced voltage, ${U}$, of the secondary winding is measured against a voltmeter, which can be written in Faraday's law as
\begin{equation}
{U}=-N_2s\frac{dB}{dt}+u_0
\label{eq4.2}
\end{equation}
where $s$ is the yoke sectional area and $u_0$ an offset in the measurement. The magnetic flux density given by (\ref{eq4.2}) can be written as
\begin{equation}
B=-\frac{1}{N_2s}\int_{T}({U}-u_0)dt
\label{eq4.3}
\end{equation}
Note that in the measurement, $u_0$ is an unknown quantity. But in practice, we can choose a constant $u_0$ value that makes the averaged $B$ field in a period ($T$) equal to zero when the excitation current has no dc component.

\subsection{Measurement results}
With the measurement of $H$ and $B$ fields, respectively presented in (\ref{eq4.1}) and (\ref{eq4.3}), the $BH$ hysteresis loop of the sample can be determined. As a comparison, two cases of measurement for the yoke material, with and without the heat treatment ($\approx$ 1150$^{\circ}$C in hydrogen for 4 hours), were made. Note that both measurements were carried out in the same yoke piece.

\begin{figure}
\center
\includegraphics[width=0.5\textwidth]{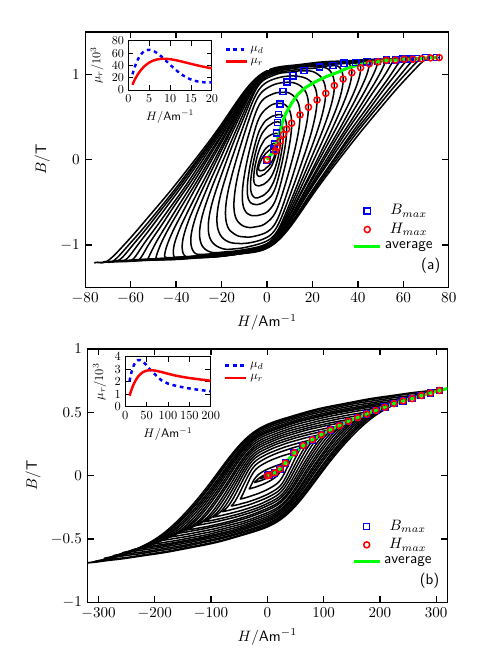}
\caption{Measurement result of the main $BH$ hysteresis loops in the yoke sample. (a) and (b) respectively show the measurement result with and without heat treatment. The main magnetization $BH$ curve is calculated by averaging the $B$-maximum and $H$-maximum points. In the subplot of each graph, the $\mu_rH$ curve and the $\mu_dH$ curve are presented.}
\label{fig42}
\end{figure}

We first measured the main $BH$ hysteresis loops. In this measurement, the signal generator supplies a sine voltage of 0.1\,Hz without the dc component. The amplitude of the excitation, i.e. $H$ field, is slowly increased to the maximum (respectively 320\,A/m and 76\,A/m before and after the material heat treatment).
The main $BH$ hysteresis loops of the yoke sample with and without heat treatment are shown in Fig. \ref{fig42} (a) and \ref{fig42} (b).  It can be seen that the hysteresis shape of the sample has a significant dependence to the heat treatment: Without the heat treatment, the main $BH$ loops have a sharp edge, where the $B$ and $H$ fields meet at the same $BH$ point and reach both maximum values. In this case, the main magnetization curve can be easily obtained by connecting these maximum field points. However, after the heat treatment, sharp edges disappear in the main $BH$ loops before reaching saturation. This is probably caused by electromagnetic resistance effects, e.g. skin effect ( For the low frequency range, the additional phase shift is proportional to $\sqrt{\hat{\mu}}$ where $\hat{\mu}$ is the average permeability of the $BH$ loop). In order to suppress the bias related to this effect, as shown in Fig. \ref{fig42}, a curve averaged by the $B$-maximum point and the $H$-maximum point is used to present the main magnetization. Accordingly, the relative permeability $\mu_r$ as a function of $H$, and the relative differential permeability $\mu_d$ as a function of $H$ can be calculated, as shown in the subplots of Fig. \ref{fig42}. Before the heat treatment, the maximum permeability is about 2900 at $H\approx60$\,A/m while the maximum $\mu_d$ is about 3700 at $H\approx30$\,A/m, while after the heat treatment, $\mu_r$ has a maximum value of 51000 at $H=8.5$\,A/m and $\mu_d$ reaches the maximum at $H=5$\,A/m. It is concluded that heat treatment improves the permeability of the yoke sample by a factor of around 17.

\begin{figure}
\center
\includegraphics[width=0.48\textwidth]{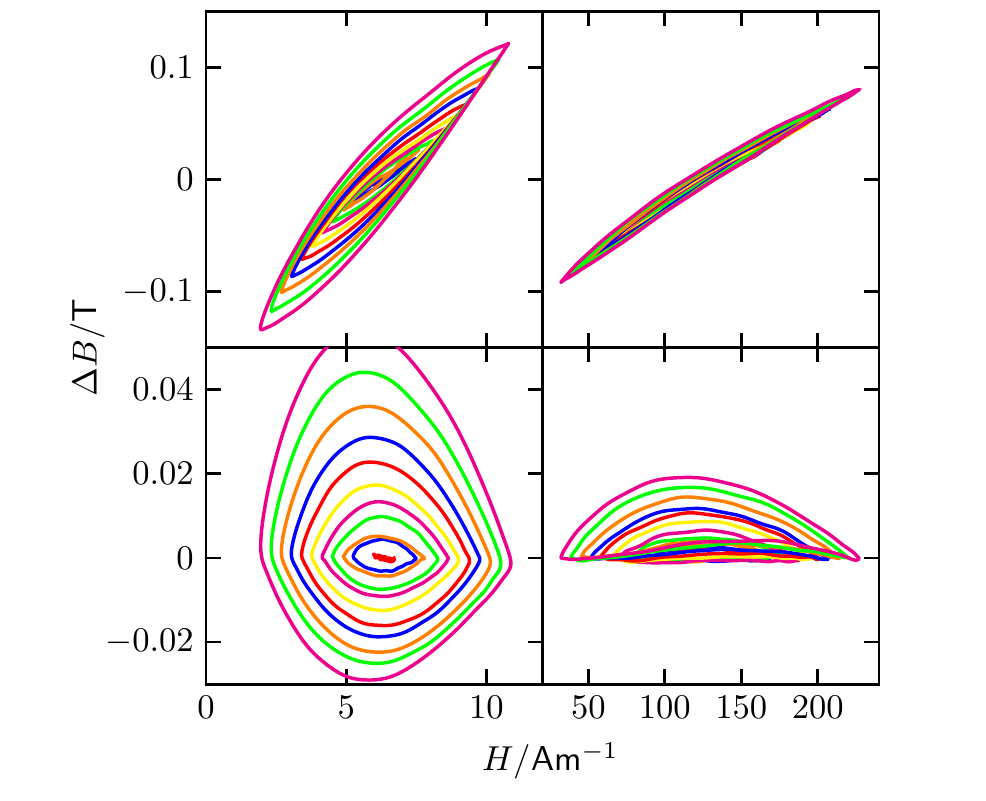}
\caption{The measurement result of minor hysteresis loops. The upper subplots are original measurement results with a linear component while the lower are normalized hysteresis loops as described in Fig. \ref{fig02}. The left and right are two independent measurements with and without heat treatment.}
\label{fig43}
\end{figure}

With the same experimental configuration, if a dc component is added to the excitation current $\mathcal{I}$, it is then allowed to measure the minor $BH$ hysteresis loop of the sample. The measurement without heat treatment was made with the $H$ field centered at 130\,A/m. The $H$ field amplitude varies from 65\,A/m to 200\,A/m by changing the ac excitation amplitude. The set point after heat treatment is at $H=$\,6.3A/m and the $H$ field changes in the range of 1\,A/m to 12\,A/m. These configurations, where the $B$ field is about 0.4\,T, are close to the real working point of a Kibble balance magnetic circuit.

The measurement result of the minor hysteresis loops is shown in the upper subplots of Fig. \ref{fig43}. As discussed in section \ref{sec02}, the linear component of the $B$ field change does not contribute to the hysteresis effect, therefore, we removed the linear component and normalized these minor hysteresis loops as shown in the lower subplots. It can be seen from the measurement result that the non-linearity of the minor $BH$ loops behavior differs in two $H$ changing directions, and the normalized loop has also a dependence on the heat treatment.

\subsection{An evaluation of the hysteresis effect}
\begin{figure}
\center
\includegraphics[width=0.5\textwidth]{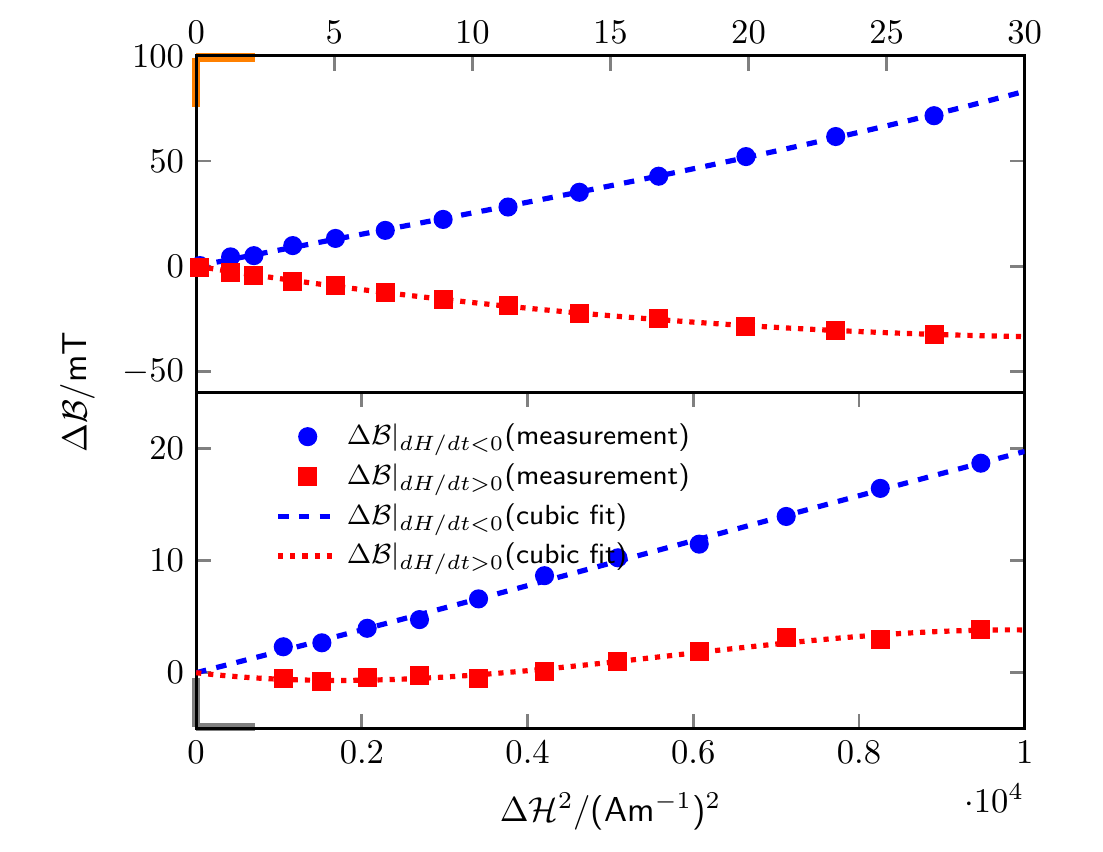}
\caption{Measurement and fit results of $\Delta \mathcal{B}(\Delta \mathcal{H}^2)$ functions in both $H$ increasing ($dH/dt>0$) and $H$ decreasing ($dH/dt<0$) directions. The upper and lower plots show the results of the yoke sample with and without the heat treatment, respectively.}
\label{fig44}
\end{figure}

Knowing the minor hysteresis loops as shown in Fig. \ref{fig43}, we can obtain the $\Delta B$ value at the $H$ field center ($\Delta H_y=0$), i.e. $\Delta\mathcal{B}$, as a function of the $H$ field change $\Delta \mathcal{H}$. Fig. \ref{fig44} presents $\Delta \mathcal{B}(\Delta \mathcal{H}^2)$ functions in $H$ increasing ($dH/dt>0$) and $H$ decreasing ($dH/dt<0$) directions.
In both directions, we use a cubic fit, i.e.
\begin{equation}
\Delta \mathcal{B}=\chi_2\Delta \mathcal{H}^2+\chi_4\Delta \mathcal{H}^4+\chi_6\Delta \mathcal{H}^6,
\label{eq4.5}
\end{equation}
to model the $\Delta \mathcal{B}(\Delta \mathcal{H}^2)$ function. It can be seen in Fig. \ref{fig44} that the fit of (\ref{eq4.5}) can well represent the measurement data. The fit is then used to interpolate the $\Delta \mathcal{B}$ value in the weighing measurement of the Kibble balance, where $\Delta \mathcal{H}$ in this case should be calculated based on the magnetic field change due to the coil current, i.e. $\Delta \mathcal{H}=\Delta B/(\mu_0\mu_r)$. Note that $\Delta B\approx 0.6$\,mT is shown in Fig. \ref{fig04} in the BIPM magnet system.

Table \ref{tab1} presents the fitting result and parameters used for evaluating the magnetic hysteresis error. It is observed from the calculation that when $\Delta \mathcal{H}$ is small, the non-linearity of (\ref{eq4.5}) is mainly contributed by the quadratic term ($>99.9\%$ in both cases). Also, the gain factor $K$ approaches 1/3 with a difference below $1\times10^{-5}$. Note that although the high order terms contribute weakly to $\Delta \mathcal{B}$ when $\Delta \mathcal{H}$ is small, they cannot be simply removed during the fit. Because the $\Delta \mathcal{H}$ value is much larger in the fit, and these higher order terms can have a significant contribution. In Table \ref{tab1}, $\Delta \mathcal{B}_1$ and $\Delta \mathcal{B}_2$ are defined as the interpolation values according to (\ref{eq4.5}) in the weighing measurement along the $H$ decreasing and increasing directions, i.e. $\Delta \mathcal{B}_1=\Delta \mathcal{B}|_{dH/dt<0}$, $\Delta \mathcal{B}_2=\Delta \mathcal{B}|_{dH/dt>0}$. It is seen from the calculation that the main contribution comes from $\Delta \mathcal{B}_1$, and $\Delta \mathcal{B}_2$ has an opposite sign with a smaller amplitude.
As shown in (\ref{eq1x}), it is reasonable to consider that the inner and outer yokes share the same $BH$ working point to simplify the calculation. Using $K\approx1/3$, the magnetic hysteresis effect presented in (\ref{eq1x}) can be written as
\begin{eqnarray}
  \frac{B_{aw}-B_{av}}{B_{av}}\approx-\frac{\Delta \mathcal{B}_{1}+\Delta \mathcal{B}_{2}}{6B_{av}}.
  \label{sim}
\end{eqnarray}

Following (\ref{sim}), the total magnetic hysteresis effect is formed by the residual of combining  $\Delta \mathcal{B}_1$ and $\Delta \mathcal{B}_2$, in which $\Delta \mathcal{B}_2$ cancels the major part of the $\Delta \mathcal{B}_1$ component. With $B_{av}=0.4\,$T, (\ref{sim}) yields a bias of $(-21.0\pm2.4)\times10^{-9}$ (with heat treatment) and $(-16.8\pm1.4)\times10^{-9}$ (without heat treatment) in the Kibble balance measurement. Note that here the uncertainty ($k=1$) is mainly from the $\Delta\mathcal{H}^2$ determination: The maximum error for  $\Delta B_y$ determination is 3.9\% obtained from figure \ref{fig04}, and the $\mu$ uncertainty is assigned by the standard deviation of the measurement, 4.2\% and 1.0\%, respectively for samples with and without heat treatment. Since the two effects are comparable, apparently, the magnetic hysteresis effect cannot be limited by the yoke heat treatment. A minus sign means that the yoke hysteresis will lower the magnetic field at the coil position in the weighing measurement. In a Kibble balance, the magnetic flux density decrease in the weighing phase will be compensated by a feedback current, which increases the realized mass in the new SI.

\begin{table}[tp!]
\caption{Evaluation results of the magnetic hysteresis error.}
\centering
\begin{tabular}{l l r r}
\hline\hline
Parameters 					& Unit 							& Before HT 		& After HT 			\\
\hline
$B_{av}, B_{yv}$            &T                              &    0.4    		&    0.4    		\\
$H_{yv}$                    &A/m                            &    130    		&    6.3    		\\
$\mu_r$                 	&rel.                           &    2400    		&    48700    		\\
$\mu_d$                 	&rel.                           &    1600    		&    62700    		\\
$\Delta B_y$              	&T                              &    0.0006    		&    0.0006    		\\
$\Delta \mathcal{H}$        &A/m                            &    1.989E-01    	&    9.804E-03    	\\
$\Delta \mathcal{H}^2$      &(A/m)$^{2}$                    &    3.958E-02    	&    9.612E-05    	\\
$\chi_{2,dH/dt<0}$        	&T/(A/m)$^{2}$                  &    1.871E-06    	&    2.578E-03    	\\
$\chi_{4,dH/dt<0}$        	&T/(A/m)$^{4}$                  &    2.271E-11    	&    -1.033E-05    	\\
$\chi_{6,dH/dt<0}$        	&T/(A/m)$^{6}$                  &    -1.246E-15    	&    5.604E-07    	\\
$\chi_{2,dH/dt>0}$        	&T/(A/m)$^{2}$                  &    -8.519E-07    	&    -2.053E-03    	\\
$\chi_{4,dH/dt>0}$        	&T/(A/m)$^{4}$                  &    2.917E-10    	&    3.300E-05    	\\
$\chi_{6,dH/dt>0}$        	&T/(A/m)$^{6}$                  &    -1.684E-14    	&    -5.996E-08    	\\
$\Delta\mathcal{B}_1$   	&T                              &    7.40E-08    	&    2.478E-07    	\\
$\Delta\mathcal{B}_2$   	&T                              &    -3.37E-08    	&    -1.973E-07    	\\
\hline
Total effect            	&$\times10^{-9}$              	&-16.8        		& -21.0				\\
Uncertainty($k=2$)          &$\times10^{-9}$                &2.8              	& 4.8				\\
\hline
\hline
HT=heat treatment
\end{tabular}
\label{tab1}
\end{table}

\section{Discussion}
\label{sec04}
The magnetic hysteresis effect could have a dependence on the chemical composition of the yoke material. In the following discussion, we do not focus on the yoke material itself, but the related parameters when the yoke magnetic property is known. Some dependence and possible optimization of the magnetic hysteresis effect are summarized as follows.

Eqs. (\ref{eq14}) and (\ref{eq15}) show that the yoke magnetic field change in the weighing measurement is proportional to the coil ampere-turns, $NI$, and inversely proportional to the air gap width $\delta$. Because the hysteresis error is a square effect of the yoke $H$ field change, a smaller $NI$ or a larger $\delta$ in the electromagnet system greatly helps to reduce the effect. Since $NI\propto mg/(2\pi r_cB_a)$, if the test mass $m$ is fixed, the hysteresis effect is then proportional to $1/(B_ar_c\delta)^2$. For example, the $1/(B_ar_c\delta)^2$ value of the NIST-4 system \cite{21} is only 1/25 of the BIPM value, and hence the hysteresis of NIST-4 Kibble balance should be much weaker ($<1\times10^{-9}$) if a similar material is used.

\begin{figure}
\center
\includegraphics[width=0.5\textwidth]{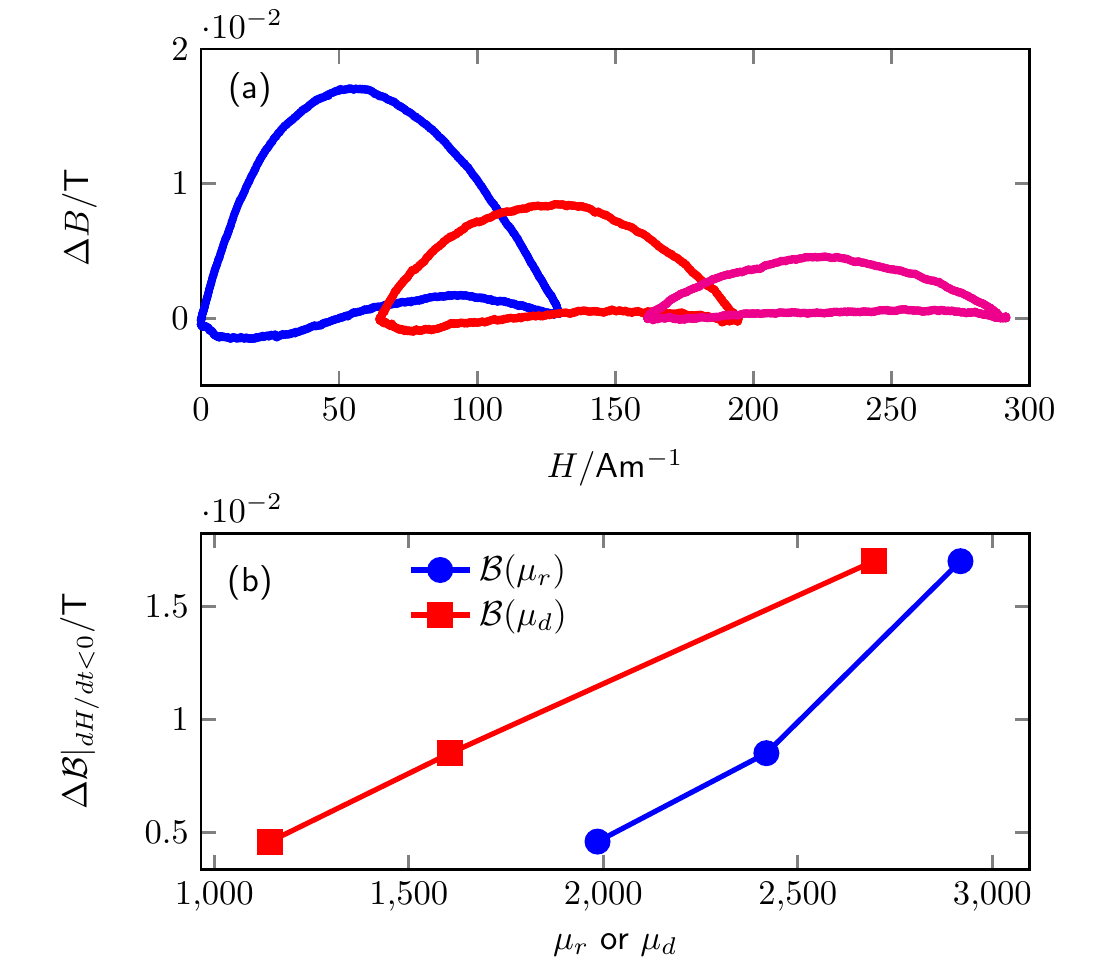}
\caption{The magnetic hysteresis dependence on the yoke permeability. (a) presents the measurement result of normalized minor loops at three different $H$ positions: $H=65$\,A/m, $H=130$\,A/m and $H=227$\,A/m. (b) shows the relationship between the peak values of three loops, i.e. $\Delta \mathcal{B}|_{dH/dt<0}$, as functions of the yoke permeability $\mu_r$ and the differential permeability $\mu_d$.} %(c) plots the ratio $\mu_d/\mu_r^2$ as a function of $H$.}
\label{fig09}
\end{figure}

The working point of the yoke, or the yoke permeability $\mu_r$ also appears in the analysis of the hysteresis effect. An interesting conclusion obtained from the result in Table \ref{tab1} is that the magnetic hysteresis error, in fact, is not very sensitive to the heat treatment, or the yoke permeability, $\mu$. This is because when $\mu$ is increased, on the one hand,  $\Delta\mathcal{H}$ becomes smaller, but on the other hand, the coefficients of the fit (mainly $\chi_2$) become larger. Also, from the measurement result, it seems that the symmetry of the normalized minor loops is better with a larger yoke permeability.

It would be also interesting to analyze the hysteresis error change when the yoke working point shifts along a fixed $BH$ curve. First, the magnetic field change in the yoke is inversely proportional to $\mu_r$ and hence the hysteresis effect is proportional to $1/\mu_r^2$. However, the sensitivity of the magnetic flux density change $\Delta\mathcal{B}$ as a function of $\Delta \mathcal{H}$ can also be related to the yoke permeability. In order to investigate this dependence, we measured the magnetic density change with a fixed $\Delta \mathcal{H}$ at three different $H$ locations (65\,A/m, 130\,A/m and 227\,A/m) of the sample without heat treatment (The sample without heat treatment provides better resolution during the measurement). The measurement result of three normalized minor loops is shown in Fig. \ref{fig09}(a). It can be seen that the magnetic flux density change has an obvious dependence on the $H$ field.
To clarify the relationship between the yoke magnetic flux density change and its permeability, we calculated the peak value of three minor loops, i.e. $\Delta \mathcal{B}|_{dH/dt<0}$, as functions of the yoke static permeability $\mu_r$ and the yoke differential permeability $\mu_d$. The results are plotted in Fig. \ref{fig09}(b). A linear relation between $\Delta \mathcal{B}|_{dH/dt<0}$ and $\mu_d$ is obtained, therefore, we conclude that the hysteresis effect is approximately proportional to $\mu_d$.
Combining the two conclusions above, the magnetic hysteresis effect is proportional to a permeability ratio $\mu_d/\mu_r^2$. It can be easily observed that close to the working yoke permeability the $\mu_d/\mu_r^2$ value is stable, and a slight change of the yoke permeability during the weighing measurement will not significantly affect the magnetic hysteresis error.

Except for the two-mode, two-phase scheme, a Kibble balance can also be operated with a one-mode, two-phase scheme \cite{11,12}, or one-mode, one-phase scheme \cite{13}. Under a one-mode measurement, the current is through the coil during both weighing and velocity measurement phases. Compared to the two-mode, two-phase scheme, the coil current change and hence the yoke magnetic status change between weighing and velocity measurements is much less (at least two magnitudes smaller), therefore, the magnetic hysteresis effect in the above examples is negligible ($<1\times10^{-9}$) and should not be a limitation in the BIPM one-mode, two-phase measurement.

\section{Conclusion}
\label{sec05}
The yoke magnetic hysteresis error is a part of the current magnetization effect which arises from the $BH$ non-linear characteristic of the yoke material. Understanding its mechanism helps to characterize the performance of a yoke-based magnetic circuit, and may also lead optimization in designing such systems. In this paper, we presented both a theoretical analysis and a practical technique for evaluating the magnetic hysteresis error based on measuring yoke minor hysteresis loops.

Theoretical analysis shows the magnetic hysteresis error is a nonlinear current effect. The yoke status change is mathematically described by a normalized minor loop. Based on this description, the yoke magnetic flux density change $\Delta B_y$ is modeled by the yoke $H$ field change $\Delta\mathcal{H}$, while $\Delta\mathcal{H}$ is linked to the coil magnetic field in the air gap following a continuous boundary condition. In this way, the hysteresis error is quantitatively related to the coil current.

Experimental measurement of a soft yoke sample has been carried out to check the proposed theory. The measurement and proposed evaluation technique showed how the hysteresis effect is related to even orders of the yoke field change caused by the coil current. An evaluation of the hysteresis effect based on this experimental determination yields an effect of about 2 parts in $10^{8}$ under a configuration of the two-mode, two-phase scheme in the BIPM system. The one-mode scheme has the advantage of suppressing the magnetic hysteresis error. As observed the effect depends closely on the coil ampere-turns, the width of the air gap, and the yoke property. As demonstrated and discussed in this paper, the non-linear current effect can be significant in Kibble balances (especially for magnet systems with a small gap), which should be checked or optimized carefully: 1) conventionally, the nonlinear magnetic effect can be determined experimentally by weighing different masses; 2) as presented in \cite{5}, with an appropriate mechanical design, the hysteresis error can be removed by ramping the weighing current slowly to zero before each velocity measurement.

In the end of the paper, we would like to acknowledge some unaddressed consequences of the magnetic hysteresis in this work. It is assumed that the yoke magnetic status is repeatable in a full measurement cycle, e.g. velocity-weighing (mass on/off), which in reality may shift with the environmental change and the repeatability of the current ramping. Besides, in the weighing measurement, it probably needs several mass on/off cycles to stabilize the magnetic state and the first measurement may differ from the ones after. The estimation assumes a weighing position at $z=0$, where the hysteresis effect, in fact, is minimum, because when the weighing position is chosen shifted from $z=0$, the $H$ field change will increase in one vertical end of the coil, and decrease on the other. The major effect (quadratic term) in this case is no longer symmetrical, which will lead to a larger average factor $K$ in the coil region.

\section*{Appendix}
In Fig. \ref{fig02}(b), the $H$ increasing curve $(dH/dt>0)$ and the $H$ decreasing curve $(dH/dt<0)$ in the original loop are respectively written in forms of polynomials, i.e.
\begin{equation}
\Delta B_y|_{dH/dt>0}=\sum_{i=0}^{\infty}\lambda_i\Delta H_y^i,
\label{a1}
\end{equation}
\begin{equation}
\Delta B_y|_{dH/dt<0}=\sum_{i=0}^{\infty}\gamma_i\Delta H_y^i,
\label{a2}
\end{equation}
where $\lambda_i$ and $\gamma_i$ are polynomial coefficients of two curves. As (\ref{a1}) goes through point V, i.e. $\Delta B_y|_{dH/dt>0}=0$ when $\Delta H=0$, and hence $\lambda_0=0$. Therefore, (\ref{a1}) can be rewritten as
\begin{equation}
\Delta B_y|_{dH/dt>0}=\sum_{i=1}^{\infty}\lambda_i\Delta H_y^i.
\label{a3}
\end{equation}
Eqs. (\ref{a2}) and (\ref{a3}) have crossing points at $\Delta H_y=\pm\Delta \mathcal{H}$, then we have
\begin{equation}
\sum_{i=1}^{\infty}\lambda_i\Delta \mathcal{H}^i=\sum_{i=0}^{\infty}\gamma_i\Delta \mathcal{H}^i,
\label{a4}
\end{equation}
\begin{equation}
\sum_{i=1}^{\infty}\lambda_i(-\Delta \mathcal{H})^i=\sum_{i=0}^{\infty}\gamma_i(-\Delta \mathcal{H})^i.
\label{a5}
\end{equation}
As is known in the analysis, the yoke magnetic flux density change due to the yoke hysteresis is $-(\overline{AV}+\overline{AC})$, which based on (\ref{a1})-(\ref{a5}) can be written as
\begin{eqnarray}
  -(\overline{AV}+\overline{AC}) &=& \left[\sum_{i=0}^{\infty}\gamma_i\Delta \mathcal{H}^i+\sum_{i=1}^{\infty}\gamma_i(-\Delta \mathcal{H})^i\right]-\lambda_0-\gamma_0\nonumber\\
   &=&\sum_{i=0,2,4,...}\gamma_i\Delta \mathcal{H}^i.
   \label{a6}
\end{eqnarray}

It can be seen from (\ref{a6}) that the yoke magnetic flux density change due to the hysteresis contains only even order terms of $\Delta \mathcal{H}$.

The normalized hysteresis curves in Fig. \ref{fig02}(b) are obtained by removing a linear component that through two end points of the original hysteresis curves. First, the line through point A and two end points of the original loop curve is solved as
\begin{eqnarray}
\Delta B_y&=&\frac{\displaystyle\sum_{i=0}^{\infty}\gamma_i\Delta \mathcal{H}^i+\displaystyle\sum_{i=0}^{\infty}\gamma_i(-\Delta \mathcal{H})^i}{2}\nonumber\\
&+&\frac{\displaystyle\sum_{i=0}^{\infty}\gamma_i\Delta \mathcal{H}^i-\displaystyle\sum_{i=0}^{\infty}\gamma_i(-\Delta \mathcal{H})^i}{2\Delta \mathcal{H}}\Delta H_y.
\label{a7}
\end{eqnarray}
Then the normalized $H$ increasing and $H$ decreasing curves can be then written as
\begin{eqnarray}
\Delta \mathcal{D}|_{dH/dt>0}&=&\Delta {B}_y|_{dH/dt>0}-\Delta B_y\nonumber\\
&=&\sum_{i=1}^{\infty}\lambda_i\Delta H_y^i-\frac{\displaystyle\sum_{i=0}^{\infty}\gamma_i\Delta \mathcal{H}^i+\displaystyle\sum_{i=0}^{\infty}\gamma_i(-\Delta \mathcal{H})^i}{2}\nonumber\\
&-&\frac{\displaystyle\sum_{i=0}^{\infty}\gamma_i\Delta \mathcal{H}^i-\displaystyle\sum_{i=0}^{\infty}\gamma_i(-\Delta \mathcal{H})^i}{2\Delta \mathcal{H}}\Delta H_y.
\label{a8}
\end{eqnarray}
\begin{eqnarray}
\Delta \mathcal{D}|_{dH/dt<0}&=&\Delta {B}_y|_{dH/dt<0}-\Delta B_y\nonumber\\
&=&\sum_{i=0}^{\infty}\gamma_i\Delta H_y^i-\frac{\displaystyle\sum_{i=0}^{\infty}\gamma_i\Delta \mathcal{H}^i+\displaystyle\sum_{i=0}^{\infty}\gamma_i(-\Delta \mathcal{H})^i}{2}\nonumber\\
&-&\frac{\displaystyle\sum_{i=0}^{\infty}\gamma_i\Delta \mathcal{H}^i-\displaystyle\sum_{i=0}^{\infty}\gamma_i(-\Delta \mathcal{H})^i}{2\Delta \mathcal{H}}\Delta H_y.
\label{a9}
\end{eqnarray}
Based on (\ref{a8}) and (\ref{a9}), the magnetic field change in the normalized hysteresis curves can be written as
\begin{eqnarray}
&&-(\overline{VD}+\overline{VE})\nonumber\\
&=&-\Delta\mathcal{D}|_{dH/dt>0}(\Delta H_y=0)-\Delta\mathcal{D}|_{dH/dt<0}(\Delta H_y=0)\nonumber\\
&=&-\Delta\mathcal{B}|_{dH/dt>0}-\Delta\mathcal{B}|_{dH/dt<0}\nonumber\\
&=&\sum_{i=0}^{\infty}\gamma_i\Delta \mathcal{H}^i+\sum_{i=0}^{\infty}\gamma_i(-\Delta \mathcal{H})^i\nonumber\\
&=&\sum_{i=0,2,4,...}\gamma_i\Delta \mathcal{H}^i.
\label{a10}
\end{eqnarray}
A comparison of (\ref{a6}) and (\ref{a10}) yields
\begin{equation}
-(\overline{AV}+\overline{AC})=-(\overline{VD}+\overline{VE})=\sum_{i=0,2,4,...}\gamma_i\Delta \mathcal{H}^i.
\end{equation}


\begin{thebibliography}{10}
\bibitem{kb76}
B. P. Kibble, ``A measurement of the gyromagnetic ratio of the proton by the strong field method," {\it Atomic Masses and Fundamental Constants 5}, pp. 545-551, Springer, 1976.

\bibitem{cgpm18}
Resolution 1, 26th General Conference on Weights and Measures (CGPM), 2018.
\url{https://www.bipm.org/utils/common/pdf/CGPM-2018/26th-CGPM-Resolutions.pdf}

\bibitem{darine16}
D. Haddad, {\it et al}, ``Bridging classical and quantum mechanics," {\it Metrologia}, vol. 53, no. 5, pp. A83-A85, 2016.

\bibitem{stephan16}
I. A. Robinson and S. Schlamminger, ``The watt or Kibble balance: a technique for implementing the new SI definition of the unit of mass" {\it Metrologia}, vol. 53, no. 5, pp. A46-A74, 2016.

\bibitem{4}
I. A. Robinson, ``Towards the redefinition of the kilogram: a measurement of the Planck constant using the NPL Mark II watt balance," {\it Metrologia}, vol. 49, no. 1, pp. 113-156, 2011.

\bibitem{5}
B. M. Wood, {\it et al}, ``A summary of the Planck constant determinations using the NRC Kibble balance," {\it Metrologia}, vol. 54, no. 3, pp. 399-409, 2017.

\bibitem{6}
D. Haddad, {\it et al}, ``Invited Article: A precise instrument to determine the Planck constant, and the future kilogram," {\it Rev. Sci. Instrum.}, vol. 87, no. 6, pp. 061301, 2016.

\bibitem{7}
H. Baumann, {\it et al}, ``Design of the new METAS watt balance experiment Mark II," {\it Metrologia}, vol. 50, no. 3, pp. 235-242, 2013.

\bibitem{8}
M. Thomas, {\it et al}, ``A determination of the Planck constant using the LNE Kibble balance in air," {\it Metrologia}, vol. 54, no. 4, pp. 468-480, 2017.

\bibitem{9}
C. M. Sutton C M and M. T. Clarkson, ``A magnet system for the MSL watt balance," {\it Metrologia}, vol. 51, no. 2, pp. S101-S109, 2014.

\bibitem{10}
D. Kim, {\it et al}, ``Design of the KRISS watt balance," {\it Metrologia}, vol. 51, no. 2, pp. S96-S100, 2014.

\bibitem{11}
H. Fang, {\it et al}, ``The BIPM Kibble Balance for the realization of the redefined kilogram," {\it 2018 Conference on Precision Electromagnetic Measurements (CPEM 2018)}, Paris, France, 2018.

\bibitem{12}
I. A. Robinson, {\it et al}, ``Developing the next generation of NPL Kibble balances,"  {\it 2018 Conference on Precision Electromagnetic Measurements (CPEM 2018)}, Paris, France, 2018.

\bibitem{13}
A. Picard, {\it et al}, ``The BIPM watt balance: improvements and developments," {\it IEEE Trans. Instrum. Meas.}, vol. 60, no. 7, pp. 2378-2386, 2011.

\bibitem{li1}
S. Li, Z. Zhang and B. Han, ``Nonlinear magnetic error evaluation of a two-mode watt balance experiment," {\it Metrologia}, vol. 50, no. 5, pp. 482-489, 2013.

\bibitem{li2}
S. Li, S. Schlamminger, J. R. Pratt, ``A nonlinearity in permanent-magnet systems used in watt balances," {\it Metrologia}, vol. 51, no. 5, pp. 394-401, 2014.

\bibitem{16}
C. A. Sanchez, {\it et al}, ``A determination of Planck's constant using the NRC watt balance," {\it Metrologia}, vol. 51, no. 4, pp. S5-S14, 2014.

\bibitem{17}
D. Haddad, {\it et al}, ``Measurement of the Planck constant at the National Institute of Standards and Technology from 2015 to 2017," {\it Metrologia}, vol. 54, no. 5, pp. 633-641, 2017.

\bibitem{18}
S. Li, {\it et al}, ``A permanent magnet system for Kibble balances," {\it Metrologia}, vol. 54, no. 5, pp. 775-783, 2017.

\bibitem{19}
S. Li, {\it et al}, ``Coil-current effect in Kibble balances: analysis, measurement, and optimization," {\it Metrologia}, vol. 55, no. 1, pp. 75-83, 2018.

\bibitem{21}
F. Seifert, {\it et al}, ``Construction, measurement, shimming, and performance of the NIST-4 magnet system," {\it IEEE Trans. Instrum. Meas.}, vol. 63, no. 12, pp. 3027-3038, 2014.

\bibitem{li16}
S. Li, {\it et al}, ``Coil motion effects in watt balances: a theoretical check," {\it Metrologia}, vol. 53, no. 2, pp. 817-828, 2016.

\bibitem{22}
S. Li, {\it et al}, ``Field analysis of a moving current-carrying coil
in OMOP Kibble balances," {\it 2018 International Applied Computational Electromagnetics Society Symposium (ACES2018)}, Denver, USA, 2018.
\end{thebibliography}
\end{document}